\documentclass[dvips,generic,twoside,noinfoline]{imsart}

\usepackage[OT1]{fontenc}
\usepackage{graphicx}
\usepackage{amssymb}
\usepackage{amsthm}
\usepackage{amsmath}
\usepackage{subfig}
\usepackage{natbib}

\newcommand{\Id}{{\bf 1}}
\newcommand{\bigoh}{\mathcal{O}}
\newcommand{\im}{\mathop{\rm Im}}
\newtheorem{observe}{Observation}
\newenvironment{observation}{\begin{observe} \rm}{\end{observe}}

\newtheorem{theorem}{Theorem}[section]

\newtheorem{remark1}[theorem]{Remark}

\newenvironment{remark}{\begin{remark1} \rm}{\end{remark1}}

\startlocaldefs
\def\phi{\varphi}
\def\epsilon{\varepsilon}
\endlocaldefs

\begin{document}

\begin{frontmatter}

\title{Computing the confidence levels for a root-mean-square test
       of goodness-of-fit, II}
\runtitle{Computing the confidence levels for a root-mean-square test
          of goodness-of-fit, II}

\author{\fnms{William} \snm{Perkins}\thanksref{t1}\ead[label=e1]{wperkins3@math.gatech.edu}}
\address{School of Mathematics\\Georgia Institute of Technology\\
         686 Cherry St.\\Atlanta, GA 30332-0160\\\printead{e1}}

\author{\fnms{Mark} \snm{Tygert}\thanksref{t2}\ead[label=e2]{tygert@aya.yale.edu}}
\address{Courant Institute of Mathematical Sciences\\NYU\\251 Mercer St.\\
         New York, NY 10012\\\printead{e2}}

\author{and\\}

\author{\fnms{Rachel} \snm{Ward}\thanksref{t3}\ead[label=e3]{rward@math.utexas.edu}}
\address{Department of Mathematics\\University of Texas\\
         1 University Station, C1200\\Austin, TX 78712\\\printead{e3}
}

\thankstext{t1}{Supported in part
by NSF Grant OISE-0730136
and an NSF Postdoctoral Research Fellowship.}
\thankstext{t2}{Supported in part
by an Alfred P. Sloan Research Fellowship.}
\thankstext{t3}{Supported in part
by an NSF Postdoctoral Research Fellowship
and a Donald D. Harrington Faculty Fellowship.}

\runauthor{W. Perkins, M. Tygert, and R. Ward}

\begin{abstract}
This paper is an extension of our earlier article,
``Computing the confidence levels for a root-mean-square test
of goodness-of-fit;'' unlike in the earlier article,
the models in the present paper involve parameter estimation ---
both the null and alternative hypotheses in the associated tests
are composite.
We provide efficient black-box algorithms
for calculating the asymptotic confidence levels of a variant
on the classic $\chi^2$ test.
In some circumstances, it is also feasible
to compute confidence levels via Monte-Carlo simulations.
\end{abstract}



\tableofcontents

\end{frontmatter}

\section{Introduction}
\label{intro}

A basic task in statistics is to ascertain whether a given set
of independent and identically distributed (i.i.d.)\ draws does not come
from any member of a specified family of probability distributions
(the specified family is known as the ``model'').
The present paper considers the case in which the draws
are discrete random variables, taking values in a finite set.
In accordance with the standard terminology,
we will refer to the possible values of the discrete random variables
as ``bins'' (``categories,'' ``cells,'' and ``classes'' are common synonyms
for ``bins'').
In earlier work, \cite{perkins-tygert-ward} treated the special case
in which the ``family'' of distributions constituting the model in fact
consists of a single, fully specified probability distribution.
The present article focuses on models parameterized with a single scalar;
our techniques extend straightforwardly to any parameterization
with multiple scalars (or, equivalently, to any parameterization
with a vector).

A natural approach to ascertaining whether a given set of i.i.d.\ draws
does not come from the model uses a root-mean-square statistic.
To construct this statistic,
we estimate both the parameter
and the probability distribution over the bins
using the given i.i.d.\ draws,
and then measure the root-mean-square
difference between this empirical distribution
and the model distribution corresponding to the estimated parameter
\citep[for details, see, for example,][page 123; or Section~\ref{simple} below]
      {rao,varadhan-levandowsky-rubin}.
If the draws do in fact arise from the specified model,
then with high probability this root-mean-square is not large.
Thus, if the root-mean-square statistic is large,
then we can be confident that the draws did not arise from the model.

To quantify ``large'' and ``confident,''
let us denote by $x$ the value of the root-mean-square
for the given i.i.d.\ draws;
let us denote by $X$ the root-mean-square statistic
constructed for different i.i.d.\ draws that definitely do in fact come
from the model.
The P-value $P$ is then defined to be the probability
that $X \ge x$ (viewing $X$ --- but not $x$ --- as a random variable).
The confidence level that the given i.i.d.\ draws do not arise
from the model is the complement of the P-value, namely $1-P$.

Unfortunately, the confidence levels for the simple root-mean-square
are different for different models.
In order to avoid this seeming inconvenience (at least asymptotically),
one may weight the average in the root-mean-square
by the reciprocals of the model probabilities associated with the various bins,
obtaining the classic~$\chi^2$ statistic
of~\cite{pearson}; see Remark~\ref{classic_remark} below.
However, with the now widespread availability of computers,
direct use of the simple root-mean-square statistic has become feasible
(and actually turns out to be very convenient).
The present paper provides efficient black-box algorithms
for computing the confidence levels for any model
with a smooth parameterization, in the limit of large numbers of draws.
Calculating confidence levels for small numbers of draws
via Monte-Carlo simulations can also be practical.
The many advantages to using the root-mean-square
have been discussed at length by~\cite{perkins-tygert-ward3}.

The remainder of the present article has the following structure:
Section~\ref{prelims} reviews previously developed techniques utilized
in the following sections.
Section~\ref{simple} details the simple statistic discussed above,
expressing the asymptotic confidence levels
for the associated goodness-of-fit test
in a form suitable for rapid computation.
Section~\ref{numerical} applies the algorithms of the present paper
to several examples.
Section~\ref{conclusion} draws some conclusions.

\section{Preliminaries}
\label{prelims}

This section summarizes a previously introduced numerical method.

The following theorem~\citep[proven in Section 3 of][]{perkins-tygert-ward}
expresses the cumulative distribution function of the sum of the squares
of independent centered Gaussian random variables 
as an integral suitable for evaluation via quadratures.

\begin{theorem}
Suppose that $n$ is a positive integer,
$X_1$,~$X_2$, \dots, $X_{n-1}$,~$X_n$ are i.i.d.\ Gaussian random variables
of zero mean and unit variance,
and $\sigma_1$,~$\sigma_2$, \dots, $\sigma_{n-1}$,~$\sigma_n$
are positive real numbers.
Suppose in addition that $X$ is the random variable
\begin{equation}
\label{summed}
X = \sum_{k=1}^n |\sigma_k \, X_k|^2.
\end{equation}

Then, the cumulative distribution function $F$ of $X$ is
\begin{equation}
\label{contoured}
F(x) = \int_0^\infty \im\left(
       \frac{e^{1-t} \, e^{it\sqrt{n}}}
       {\pi \, \bigl( t - \frac{1}{1-i\sqrt{n}} \bigr)
        \prod_{k = 1}^n \sqrt{1-2(t-1)\sigma_k^2/x+2it\sigma_k^2\sqrt{n}/x}}
       \right) \, dt
\end{equation}
for any positive real number $x$,
and $F(x) = 0$ for any nonpositive real number $x$.
The square roots in~(\ref{contoured}) denote the principal branch,
and $\,\im$ takes the imaginary part.
\end{theorem}

\pagebreak

\begin{remark}
The absolute value of the expression under the square root
in~(\ref{contoured}) is always greater than $\sqrt{n/(n+1)}$.
Therefore,
\begin{equation}
\left| \prod_{k = 1}^n
       \sqrt{1 - 2(t-1)\sigma_k^2/x + 2it\sigma_k^2\sqrt{n}/x} \right|
> \left(\frac{n}{n+1}\right)^{n/4} > \frac{1}{e^{1/4}}
\end{equation}
for any $t \in (0,\infty)$ and any $x \in (0,\infty)$.
Thus, the integrand in~(\ref{contoured}) is never large for $t \in (0,\infty)$.
\end{remark}

\begin{remark}
\label{quadratures}
An efficient means of evaluating~(\ref{contoured}) numerically
is to use adaptive Gaussian quadratures
\citep[see, for example, Section~4.7 of][]
      {press-teukolsky-vetterling-flannery}.
To attain double-precision accuracy (roughly 15-digit precision),
the domain of integration for $t$ in~(\ref{contoured}) need be only $(0,40)$
rather than the whole $(0,\infty)$.
Good choices for the lowest orders of the quadratures used in the adaptive
Gaussian quadratures are 10 and 21, for double-precision accuracy.
\citep[See Section~3 of][for the details.]{perkins-tygert-ward}
\end{remark}

\section{The simple statistic}
\label{simple}

This section details the simple root-mean-square statistic discussed briefly
in Section~\ref{intro}, determining its probability distribution
in the limit of large numbers of draws, assuming that the draws do in fact
come from the specified model.
The distribution determined in this section yields the confidence levels
(in the limit of large numbers of draws):
Given a value~$x$ for the root-mean-square statistic constructed
from i.i.d.\ draws coming from an unknown distribution,
and given the value of the maximum-likelihood estimate $\hat\theta$
for the parameter of the distribution,
the confidence level that the draws do not come from the specified model is
the probability that the root-mean-square statistic is less than $x$
when constructed from i.i.d.\ draws that do come from the model distribution
associated with the parameter~$\hat\theta$.
(Please note that the definition in~(\ref{scaled})
and~(\ref{statistic}) below of the simple statistic involves
the maximum-likelihood estimate~$\hat\theta$. Maximum likelihood
is the canonical method for parameter estimation, and is the focus
of the present paper. See formulae~(\ref{likelihood}) and~(\ref{mle}) below
regarding likelihood and maximum-likelihood estimation.)

\subsection{The distribution of the goodness-of-fit statistic}
\label{distribution}

To begin, we set notation and form the goodness-of-fit statistic $X$
to be analyzed. Given $n$ bins, numbered $1$,~$2$, \dots, $n-1$,~$n$,
we denote by $p_1(\theta)$,~$p_2(\theta)$, \dots,
$p_{n-1}(\theta)$,~$p_n(\theta)$ the probabilities associated
with the respective bins under the specified model,
where $\theta$ is a real number parameterizing the model; of course,
\begin{equation}
\label{prob}
\sum_{k=1}^n p_k(\theta) = 1
\end{equation}
for any parameter $\theta$.
In order to obtain a draw conforming to the model for a particular value
of $\theta$, we select at random one of the $n$ bins,
with probabilities $p_1(\theta)$,~$p_2(\theta)$, \dots,
$p_{n-1}(\theta)$,~$p_n(\theta)$.
We perform this selection independently $m$ times.
For $k = 1$,~$2$, \dots, $n-1$,~$n$,
we denote by $Y_k$ the fraction of times that we choose bin~$k$
(that is, $Y_k$ is the number of times that we choose bin~$k$, divided by $m$);
obviously, $\sum_{k=1}^n Y_k = 1$.
We define $X_k$ to be $\sqrt{m}$ times the difference of $Y_k$
from its expected value using the maximum-likelihood estimate $\hat\theta$
of the actual parameter $\theta$, that is,
\begin{equation}
\label{scaled}
X_k = \sqrt{m} \, (Y_k - p_k(\hat\theta))
\end{equation}
for $k = 1$,~$2$, \dots, $n-1$,~$n$.
Finally, we form the statistic
\begin{equation}
\label{statistic}
X = \sum_{k=1}^n X_k^2,
\end{equation}
and now determine its distribution in the limit that the number $m$ of draws
is large.
(The root-mean-square statistic
$\sqrt{ \sum_{k=1}^n (m Y_k - m p_k(\hat\theta))^2 / m }$
is the square root of $X$.
As the square root is a monotonically increasing function,
the confidence levels are the same whether determined via $X$
or via $\sqrt{X}$; for convenience, we focus on $X$ below.)

\begin{remark}
\label{classic_remark}
The classic $\chi^2$ test for goodness-of-fit of~\cite{pearson}
replaces~(\ref{statistic}) with the statistic
\begin{equation}
\label{classic}
\chi^2 = \sum_{k=1}^n \frac{X_k^2}{p_k(\hat\theta)},
\end{equation}
where $X_1$,~$X_2$, \dots, $X_{n-1}$,~$X_n$ are the same as
in~(\ref{scaled}) and~(\ref{statistic}),
and $\hat\theta$ is the maximum-likelihood estimate of the parameter.
\end{remark}

For definiteness, we will be assuming that
$p_1$,~$p_2$, \dots, $p_{n-1}$,~$p_n$
are differentiable as functions of the parameter $\theta$,
that the maximum of the likelihood occurs in the interior
of the domain for~$\theta$, that the maximum-likelihood estimate $\hat\theta$
is almost surely the correct value for the actual parameter $\theta$
as $m \to \infty$, and that the variance of $\hat\theta$ tends to zero
as $m \to \infty$
(thus $\hat\theta$ is not ``random'' in the limit of large numbers of draws).
As detailed, for example, by~\cite{moore-spruill}
and by~\cite{kendall-stuart-ord-arnold} in a chapter on goodness-of-fit
(see also Remark~\ref{heuristic} below),
the multivariate central limit theorem then shows that
the joint distribution of $X_1$,~$X_2$, \dots, $X_{n-1}$,~$X_n$
converges in distribution as $m \to \infty$,
with the limiting generalized probability density proportional to
\begin{equation}
\label{genden}
\exp\left(-\sum_{k=1}^n \frac{x_k^2}{2p_k(\hat\theta)} \right)
\;\cdot\; \delta\left(\sum_{k=1}^n x_k\right)
\;\cdot\; \delta\left(\sum_{k=1}^n x_k
       \, \frac{d}{d\theta} \ln(p_k(\theta))\bigg|_{\theta=\hat\theta}\right),
\end{equation}
where $\delta$ is the Dirac delta,
and $\hat\theta$ is the maximum-likelihood estimate of the parameter.

The generalized probability density in~(\ref{genden})
is a centered multivariate Gaussian distribution concentrated
on the intersection of two hyperplanes that both pass through the origin
(the intersection of the hyperplanes consists of all the points
such that $\sum_{k=1}^n x_k = 0$ and
$\sum_{k=1}^n x_k \, \frac{d}{d\theta} \ln(p_k(\theta))\big|_{\theta=\hat\theta}
= 0$);
the restriction of the generalized probability density~(\ref{genden})
to the intersection of the hyperplanes is also
a centered multivariate Gaussian.
Thus, the distribution of $X$ defined in~(\ref{statistic}) converges
as $m \to \infty$ to the distribution of the sum of the squares
of $n-2$ independent Gaussian random variables of mean zero whose variances are
the variances of the restricted multivariate Gaussian distribution
along its principal axes
\citep[see, for example, Chapter~25 of][]{kendall-stuart-ord-arnold}.
Given these variances, Remark~\ref{quadratures}
describes an efficient algorithm
for computing the probability that the associated sum of squares is less than
any particular value; this probability is the desired confidence level,
in the limit of large numbers of draws.
For a detailed discussion, see Section~\ref{explicit} below.

To compute the variances of the restricted multivariate Gaussian distribution
along its principal axes, we perform the following four steps:
\begin{enumerate}
\item Form an $n \times 2$ matrix $H$ whose columns both include
a vector that is normal to the hyperplane consisting of the points
$(x_1, x_2, \dots, x_{n-1}, x_n)$ such that
\begin{equation}
\label{hyper1}
\sum_{k=1}^n x_k = 0,
\end{equation}
and also include a vector that is normal to the hyperplane consisting
of the points $(x_1, x_2, \dots, x_{n-1}, x_n)$ such that
\begin{equation}
\label{hyper2}
\sum_{k=1}^n x_k \, \frac{d}{d\theta}\ln(p_k(\theta))\bigg|_{\theta=\hat\theta}
= 0,
\end{equation}
where $\hat\theta$ is the maximum-likelihood estimate of the parameter.
For example, we can take the entries of $H$ to be
\begin{equation}
H_{k,j} = \left\{ \begin{array}{ll}
                  1, & j = 1 \\
                  \frac{d}{d\theta}\ln(p_k(\theta))\big|_{\theta=\hat\theta},
                     & j = 2
                  \end{array} \right.
\end{equation}
for $k = 1$,~$2$, \dots, $n-1$,~$n$ and $j = 1$,~$2$,
where again $\hat\theta$ is the maximum-likelihood estimate of the parameter.
\item Form an orthonormal basis for the column space of $H$,
by constructing a pivoted $QR$ decomposition
\begin{equation}
H_{n \times 2} = Q_{n \times 2} \cdot R_{2 \times 2} \cdot \Pi_{2 \times 2},
\end{equation}
where the columns of $Q$ are orthonormal, $R$ is upper-triangular,
and $\Pi$ is a permutation matrix.
\citep[See, for example, Chapter~5 of]
      [for details on the construction of such a pivoted $QR$ decomposition.]
      {golub-van_loan}
\item Form the $n \times n$ diagonal matrix $D$ with the entries
\begin{equation}
D_{j,k} = \left\{ \begin{array}{ll}
                  1/p_k(\hat\theta), & j = k \\
                  0, & j \ne k
                  \end{array} \right.
\end{equation}
for $j,k = 1$,~$2$, \dots, $n-1$,~$n$,
where $\hat\theta$ is the maximum-likelihood estimate of the parameter.
Then, multiply $D$ from both the left and the right
by the orthogonal projection $(\Id - QQ^\top)$
onto the intersection of the hyperplanes
consisting of the points satisfying~(\ref{hyper1}) and~(\ref{hyper2}),
obtaining the $n \times n$ matrix
\begin{equation}
\label{diagonalizer}
B = (\Id - QQ^\top) \, D \, (\Id - QQ^\top),
\end{equation}
where $\Id$ is the $n \times n$ identity matrix.
\item Find the eigenvalues of the self-adjoint matrix $B$ defined
in~(\ref{diagonalizer}).
By construction, exactly two of the eigenvalues of $B$ are zeros.
The other eigenvalues of $B$ are the reciprocals of the desired variances
of the restricted multivariate Gaussian distribution along its principal axes.
\end{enumerate}

\begin{remark}
\label{faster}
The $n \times n$ matrix $B$ defined in~(\ref{diagonalizer}) is the sum
of a diagonal matrix and a low-rank matrix.
The methods of~\cite{gu-eisenstat94,gu-eisenstat95}
for computing the eigenvalues of such a matrix~$B$ require
only either $\bigoh(n^2)$ or $\bigoh(n)$ floating-point operations.
Note that the $\bigoh(n^2)$ methods of~\cite{gu-eisenstat94,gu-eisenstat95}
are more efficient than the $\bigoh(n)$ procedure of~\cite{gu-eisenstat95},
unless $n$ is impractically large.
\end{remark}

\begin{remark}
\label{heuristic}
Under appropriate regularity conditions,
it is easy to derive the homogeneous linear constraint
--- analogous to~(\ref{hyper2}) --- that
\begin{equation}
\label{constraint}
\sum_{k=1}^n X_k \, \frac{d}{d\theta}
\ln(p_k(\theta))\bigg|_{\theta=\hat\theta} = 0,
\end{equation}
where $\hat\theta$ is the maximum-likelihood estimator.
The following is a sketch of the proof of~(\ref{constraint}).

To determine the maximum-likelihood estimate $\hat\theta$,
we consider the likelihood, namely the multinomial distribution
\begin{equation}
\label{likelihood}
L(y_1,y_2,\dots,y_{n-1},y_n,\theta)
= m! \, \prod_{k=1}^n \frac{(p_k(\theta))^{my_k}}{(my_k)!}.
\end{equation}
Maximizing~(\ref{likelihood}) defines~$\hat\theta$ via the formula
\begin{equation}
\label{mle}
0 = \frac{\partial}{\partial\theta}
    \ln(L(Y_1,Y_2,\dots,Y_{n-1},Y_n,\theta))\bigg|_{\theta=\hat\theta}
  = \sum_{k=1}^n mY_k
    \, \frac{d}{d\theta} \ln(p_k(\theta))\bigg|_{\theta=\hat\theta}.
\end{equation}

It follows from~(\ref{prob}) that
\begin{equation}
\label{prob2}
\sum_{k=1}^n \frac{d}{d\theta} \, p_k(\theta) = 0
\end{equation}
for any parameter $\theta$, in particular for $\theta = \hat\theta$.
Combining (\ref{mle}) and~(\ref{prob2}) yields that
\begin{equation}
\label{almost}
\sum_{k=1}^n (Y_k - p_k(\hat\theta))
          \, \frac{d}{d\theta} \ln(p_k(\theta))\bigg|_{\theta=\hat\theta} = 0.
\end{equation}
Combining~(\ref{almost}) and~(\ref{scaled}) yields~(\ref{constraint}),
as desired.
\end{remark}

\subsection{A procedure for computing the confidence levels}
\label{explicit}
An efficient method for calculating the confidence levels
in the limit of large numbers of draws proceeds as follows.
Given i.i.d.\ draws from any distribution --- not necessarily from the model
--- we can form the associated statistic $X$ defined in~(\ref{statistic})
and~(\ref{scaled}); in the limit of large numbers of draws,
the confidence level that the draws do not arise from the model
is then just the cumulative distribution function $F(x)$
in~(\ref{contoured}) evaluated at $x = X$,
with $\sigma_k^2$ in~(\ref{contoured}) obtained via Step~4
of the algorithm of Section~\ref{distribution}
(after all, $F(x)$ is the probability that $x$ is greater than
the sum of the squares of independent centered Gaussian random variables
whose variances are given by Step~4 above).
Remark~\ref{quadratures} describes an efficient means
of evaluating $F(x)$ numerically.

\section{Numerical examples}
\label{numerical}

This section illustrates the performance of the algorithm
of Section~\ref{explicit} via several numerical examples.

\begin{figure}[p]
\begin{center}
\subfloat[]{
\rotatebox{-90}{\scalebox{.75}{\includegraphics{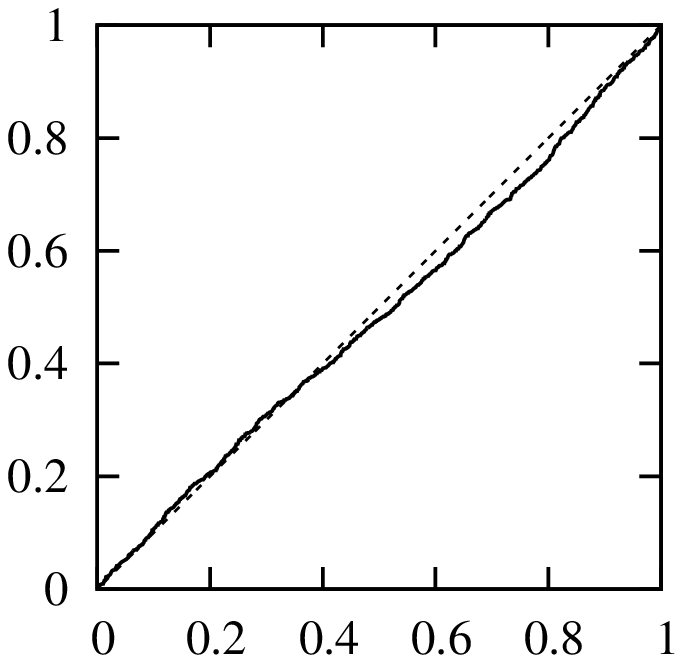}}}
\quad\quad}
\\
\subfloat[]{
\rotatebox{-90}{\scalebox{.75}{\includegraphics{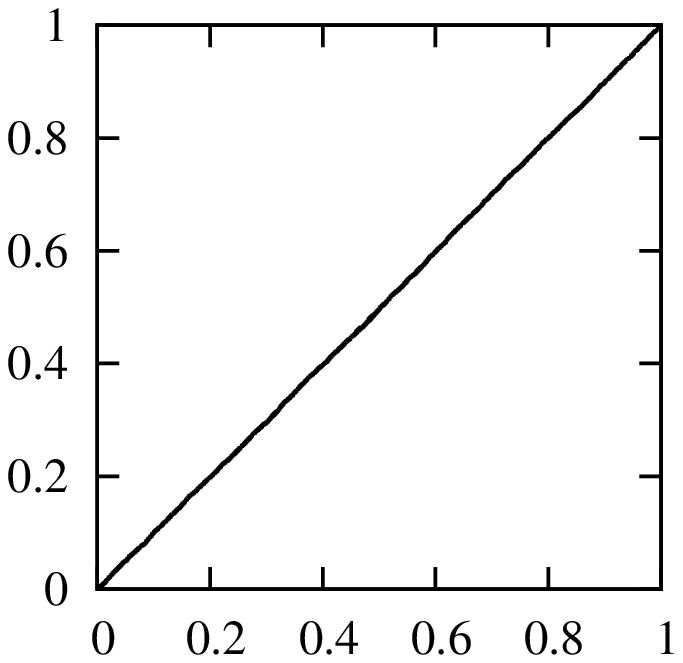}}}
\quad\quad}
\caption{$2 \times 2$ contingency-table/cross-tabulation of Table~\ref{bigtab}}
\label{conttab}
\end{center}
\end{figure}

Figure~\ref{conttab} and Table~\ref{conttabt} correspond to the first example.
The model distribution for the first example has 4 bins,
with the probabilities indicated in Table~\ref{bigtab}.
We will detail the interpretation of the figures and tables shortly.

\begin{figure}[p]
\begin{center}
\subfloat[]{
\rotatebox{-90}{\scalebox{.75}{\includegraphics{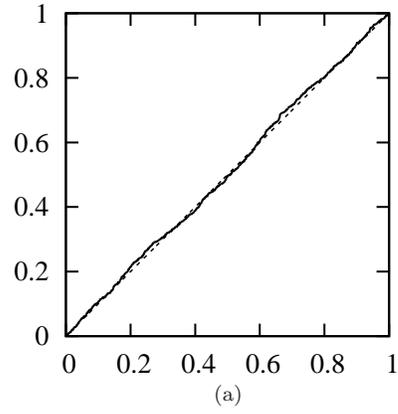}}}
\quad\quad}
\\
\subfloat[]{
\rotatebox{-90}{\scalebox{.75}{\includegraphics{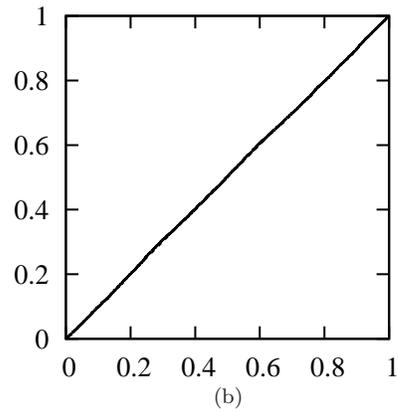}}}
\quad\quad}
\caption{Zipf distribution of Table~\ref{bigtab}}
\label{zipf}
\end{center}
\end{figure}

Figure~\ref{zipf} and Table~\ref{zipft} correspond to the second example.
The model for the second example is the Zipf distribution on 100 bins.
The row for Figure/Table~\ref{zipf} in Table~\ref{bigtab}
provides a definition of the Zipf distribution.

\begin{figure}[p]
\begin{center}
\subfloat[]{
\rotatebox{-90}{\scalebox{.75}{\includegraphics{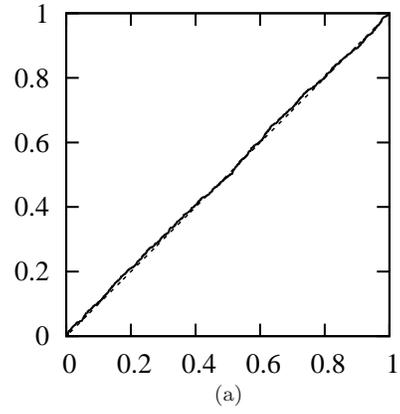}}}
\quad\quad}
\\
\subfloat[]{
\rotatebox{-90}{\scalebox{.75}{\includegraphics{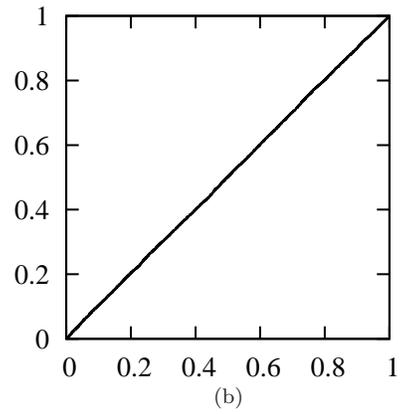}}}
\quad\quad}
\caption{Poisson distribution of Table~\ref{bigtab}}
\label{poisson}
\end{center}
\end{figure}

Figure~\ref{poisson} and Table~\ref{poissont} correspond to the third example.
The model for the third example is the standard Poisson distribution.
The row for Figure/Table~\ref{poisson} in Table~\ref{bigtab}
provides a definition of the Poisson distribution.

To test our algorithms, we conduct computational simulations.
In every simulation, we choose the number $m$ of draws
to be a very large number, namely $m =$~100,000.
(The algorithms of the present paper concern the limit as $m \to \infty$.)
Part~(a) of the examples uses $j =$~1,000 simulations;
part~(b) of the examples uses $j =$~10,000 simulations.
The convergence (as $j$ increases) of the plotted points
to the straight line of unit slope through the origin provides
numerical validation of our algorithms, for the following reasons.

To create the plots, we run $j$ simulations,
each taking $m =$~100,000 i.i.d.\ draws
from the model distribution with the specified parameter $\theta$.
For each simulation, we compute the statistic $X$ defined in~(\ref{statistic}),
forming $Y_1$,~$Y_2$, \dots, $Y_{n-1}$,~$Y_n$
and $\hat\theta$ needed in~(\ref{scaled}) and~(\ref{statistic})
using the generated draws.
We then compute the asymptotic confidence level associated with each
of these values for $X$, as described in Section~\ref{explicit},
and sort the resulting confidence levels.
These sorted results are the vertical coordinates of the points in the plots;
the horizontal coordinates are the equispaced numbers
$1/(2j)$, $3/(2j)$, \dots, $(2j-3)/(2j)$,~$(2j-1)/(2j)$.

As the number $j$ of simulations increases, and insofar as the number $m$
of draws is very large, the plotted points should converge
to the straight line through the origin of slope 1 (and, indeed,
our experiments demonstrate this).
The dotted line in each plot is the straight line through the origin
of slope 1.
The trials converge correctly:
The root-mean-square statistics for about $\alpha\%$ of the simulations
should have P-values of $\alpha\%$ or less,
for every $\alpha \in (0,100)$;
in the limit that both the number $m$ of draws
and the number $j$ of simulations are large,
the computed P-values for exactly $\alpha\%$ of the simulations
should be less than or equal to $\alpha\%$
(this follows from the definition of P-values;
it also follows from the fact that the confidence levels
for the statistic $X$ are given by its cumulative distribution function $F$,
and from the fact that $F(X)$ is uniformly distributed over $(0,1)$
for any random variable $X$ distributed according
to a continuous cumulative distribution function $F$).

The following list describes the headings of the tables:
\begin{itemize}
\item $j$ is the number of simulations conducted in generating
the associated plot.
\item $\theta$ is the parameter for the model distribution
used in generating the i.i.d.\ draws.
\item $n$ is the number of bins/categories/cells/classes in the model
(see Remark~\ref{truncator} regarding the Poisson distribution
of the third example).
\item $l$ is the maximum number of quadrature nodes
required to evaluate the confidence level
for any of the $j$ root-mean-square statistics produced by the simulations.
\item $t$ is the total number of seconds required to perform the quadratures
for evaluating the confidence levels
for all $j$ of the root-mean-square statistics produced by the simulations.
\item $s$ is the total number of seconds required to perform
all $j$ simulations.
\item $p_k(\theta)$ is the probability associated with bin $k$
($k = 1$, $2$, \dots, $n-1$, $n$), as a function of the parameter $\theta$.
\item $\hat\theta(Y_1, Y_2, \dots, Y_{n-1}, Y_n)$
is the maximum-likelihood estimate of the parameter $\theta$,
as a function of the fractions $Y_1$,~$Y_2$, \dots, $Y_{n-1}$,~$Y_n$
of the draws in the respective bins
(Section~\ref{simple} provides a detailed definition
of $Y_1$,~$Y_2$, \dots, $Y_{n-1}$,~$Y_n$).
\end{itemize}

We used Fortran~77 and ran all examples on one core
of a 2.2~GHz Intel Core~2 Duo microprocessor with 2~MB of L2 cache.
Our code is compliant with the IEEE double-precision standard
(so that the mantissas of variables have approximately one bit of precision
less than 16 digits, yielding a relative precision of about 2E--16).
We diagonalized the matrix $B$ defined in~(\ref{diagonalizer})
using the Jacobi algorithm
\citep[see, for example, Chapter~8 of][]{golub-van_loan},
not taking advantage of Remark~\ref{faster}.
We generated the pseudorandom numbers used in the simulations
via (Mitchell-Moore-Brent-Knuth) lagged Fibonacci sequences
\citep[see, for example, Section~7.1.5 of][]
      {press-teukolsky-vetterling-flannery}.

\begin{observation}
\label{infinitely}
It is easy to compute the confidence levels
(in the limit of large numbers of draws)
for a distribution having infinitely many bins,
but only to any arbitrary accuracy that is greater than the machine precision.
Specifically, given a fully specified model distribution
and an extremely small positive real number $\epsilon$,
we would retain the smallest possible number of bins
whose associated probabilities
$p_1$,~$p_2$, \dots, $p_{n-1}$,~$p_n$
satisfy $p_1 + p_2 + \dots + p_{n-1} + p_n \ge 1-\epsilon$,
and then proceed with the computation
as if these finitely many were the only bins.
When there is a parameter $\theta$ being estimated,
we observe that the maximum-likelihood estimate $\hat\theta$ typically
has variance zero and is almost surely correct in the limit of large numbers
of draws; thus, as before, we may retain the smallest possible number of bins
whose associated probabilities
$p_1(\hat\theta)$,~$p_2(\hat\theta)$, \dots,
$p_{n-1}(\hat\theta)$,~$p_n(\hat\theta)$ satisfy
$p_1(\hat\theta) + p_2(\hat\theta) + \dots
+ p_{n-1}(\hat\theta) + p_n(\hat\theta)
\ge 1-\epsilon$, and then proceed with the computation
as if these finitely many were the only bins.
(Needless to say, if the fraction of the experimental draws
falling outside the finitely many retained bins is significantly greater
than $\epsilon$, then we can be highly confident that the draws did not arise
from the model.)
\end{observation}

\begin{remark}
For the second example (the Zipf distribution),
we computed the maximum-likelihood estimate $\hat\theta$
from the data $Y_1$,~$Y_2$, \dots, $Y_{n-1}$,~$Y_n$ by finding the zero of
the function $g(\hat\theta) = f(\hat\theta) - \sum_{k=1}^n Y_k \, \ln(k) = 0$,
where $f$ is the same as in Table~\ref{bigtab}, namely
$f(\hat\theta) = \left(\sum_{k=1}^n k^{-\hat\theta} \, \ln(k)\right)
   \!\Big/\! \left(\sum_{k=1}^n k^{-\hat\theta}\right)$.
We evaluated the zero~$\hat\theta$ numerically, via bisection
\citep[see, for example, Chapter~9 of][]{press-teukolsky-vetterling-flannery}.
\end{remark}

\begin{remark}
\label{truncator}
For the third example (the Poisson distribution),
we employed Observation~\ref{infinitely}, with $\epsilon = 10^{-8}$.
\end{remark}

\begin{table}[p]
\begin{center}
\vspace{.4in}
\caption{Values for Figure~\ref{conttab}}
\label{conttabt}
\begin{tabular}{ccccccc}
\\
    &    $j$ & $\theta$ & $n$ & $l$ &   $t$ &   $s$ \\\hline
(a) & $10^3$ &      .03 &   4 & 190 & .43E0 & .44E1 \\
(b) & $10^4$ &      .03 &   4 & 190 & .43E1 & .45E2
\end{tabular}
\end{center}
\end{table}

\begin{table}[p]
\vspace{.4in}
\begin{center}
\caption{Values for Figure~\ref{zipf}}
\label{zipft}
\begin{tabular}{ccccccc}
\\
    &    $j$ & $\theta$ & $n$ & $l$ &   $t$ &   $s$ \\\hline
(a) & $10^3$ &        1 & 100 & 350 & .92E1 & .13E2 \\
(b) & $10^4$ &        1 & 100 & 390 & .11E3 & .13E3
\end{tabular}
\end{center}
\end{table}

\begin{table}[p]
\begin{center}
\vspace{.4in}
\caption{Values for Figure~\ref{poisson}}
\label{poissont}
\begin{tabular}{ccccccc}
\\
    &    $j$ & $\theta$ & $n$ & $l$ &   $t$ &   $s$ \\\hline
(a) & $10^3$ &     10.3 &  36 & 290 & .37E1 & .86E1 \\
(b) & $10^4$ &     10.3 &  36 & 330 & .37E2 & .86E2
\end{tabular}
\end{center}
\end{table}

\begin{table*}
\begin{center}
\caption{Values for Figures~\ref{conttab}--\ref{poisson}
                and Tables~\ref{conttab}--\ref{poisson}}
\label{bigtab}
\vspace{-.05in}
\begin{tabular}{ccc}
\\\\
    $\,$Fig./Table$\,$\# & $p_k(\theta)$
  & $\hat\theta(Y_1, Y_2, \dots, Y_{n-1}, Y_n)$\\\hline\\\\
Fig./Table~\ref{conttab} &
\parbox{1.3in}{$p_1 = .04 \cdot \theta$,\ $p_2 = .04 (1-\theta)$\\
               $p_3 = .96 \cdot \theta$,\ $p_4 = .96 (1-\theta)$}
                           & $\hat\theta = Y_1 + Y_3$ \\\\
     Fig./Table~\ref{zipf} & $p_k = k^{-\theta}\big/\sum_{i=1}^n i^{-\theta}$ &
\parbox{2.2in}{\begin{center}
                $\hat\theta = f^{-1}\bigl(\sum_{k=1}^n Y_k \, \ln(k)\bigr),$\\
$f(\hat\theta) = \left(\sum_{k=1}^n k^{-\hat\theta} \, \ln(k)\right)
                   \!\Big/\! \left(\sum_{k=1}^n k^{-\hat\theta}\right)$
                \end{center}} \\\\
Fig./Table~\ref{poisson} & $p_k = e^{-\theta} \theta^{k-1}/(k-1)!$
                         & $\hat\theta = \sum_{k=1}^{\infty} (k-1) \, Y_k$
\end{tabular}
\end{center}
\vspace{-.05in}
\end{table*}

\section{Conclusion}
\label{conclusion}

This paper provides efficient black-box algorithms
for computing the confidence levels
for one of the simplest, most natural goodness-of-fit statistics,
in the limit of large numbers of draws.
Although the present paper focuses on families of probability distributions
parameterized with a single scalar (and the predecessor to this article
focuses on fully specified distributions), our methods extend straightforwardly
to any parameterization with multiple scalars
(or, equivalently, to any parameterization with a vector).
Furthermore, our methods can handle arbitrarily weighted means
in the root-mean-square, in addition to the usual, uniformly weighted average
considered above.

There are many advantages to using the simple root-mean-square,
as shown by~\cite{perkins-tygert-ward3}.
With the now widespread availability of computers,
calculating the relevant P-values via Monte-Carlo simulations
can be feasible; the algorithms of the present paper can also be suitable,
and are efficient and easy-to-use.

\section*{Acknowledgements}

We would like to thank Tony Cai, Jianqing Fan, Peter W. Jones, Ron Peled,
and Vladimir Rokhlin for many helpful discussions.

\bibliographystyle{imsart-nameyear.bst}
\bibliography{stat}

\end{document}